\documentclass{nature}
\usepackage{color}
\usepackage{graphicx}
\usepackage{indentfirst}
\usepackage{caption}
\usepackage{amsmath}
\title{Terahertz semiconductor laser chaos}

\author{Binbin Liu$^{1,2,9}$, Carlo Silvestri$^{3,9}$, Kang Zhou$^{1,5,9}$, Xuhong Ma$^{1,2}$, Shumin Wu$^{1,2}$, Ziping Li$^{1}$, Wenjian Wan$^{1}$, Zhenzhen Zhang$^{1}$, Ying Zhang$^{4}$, Junsong Peng$^{4}$, Heping Zeng$^{4,5\star}$, Cheng Wang$^{6}$, Massimo Brambilla$^{7}$, Lorenzo Columbo$^{8,\star}$, and Hua Li$^{1,2\star}$}

\begin{document}

\linespread{1.5}\selectfont

\maketitle

\begin{affiliations}
\item State Key Laboratory of Materials for Integrated Circuits and Key Laboratory of Terahertz Solid State Technology, Shanghai Institute of Microsystem and Information Technology, Chinese Academy of Sciences, 865 Changning Road, Shanghai 200050, China.
 \item Center of Materials Science and Optoelectronics Engineering, University of Chinese Academy of Sciences, Beijing 100049, China.
 \item Institute of Photonics and Optical Science (IPOS), School of Physics, The University of Sydney, NSW 2006, Australia
 \item State Key Laboratory of Precision Spectroscopy, East China Normal University, Shanghai 200241, China.
 \item Chongqing Key Laboratory of Precision Optics, Chongqing Institute of East China Normal University, Chongqing 401120, China.
 \item School of Information Science and Technology, ShanghaiTech University, 393 Middle Huaxia Road, Shanghai 201210, China. 
 \item Dipartimento Interateneo di Fisica, Politecnico di Bari e CNR-IFN (UOS Bari), Via Amendola 173, IT-70126, Italy. 
 \item Dipartimento di Elettronica e Telecomunicazioni, Politecnico di Torino, Corso Duca degli Abruzzi 24, Torino, IT-10129, Italy.
 \item These authors contributed equally. \\
  $^{\star}$Corresponding author. E-mail: hpzeng@phy.ecnu.edu.cn; lorenzo.columbo@polito.it; hua.li@mail.sim.ac.cn.
\end{affiliations}

\begin{abstract}
Chaos characterized by its irregularity and high sensitivity to initial conditions finds various applications in secure optical communications, random number generations, light detection and ranging systems, etc. Semiconductor lasers serve as ideal light platforms for chaos generations owing to the advantages in on-chip integration and complex nonlinear effects. In near-infrared wavelengths, semiconductor laser based chaotic light sources have been extensively studied and experimentally demonstrated. However, in the terahertz (THz) spectral range, due to the lack of effective THz light sources and high-speed detectors, chaos generation in THz semiconductor lasers, e.g., quantum cascade lasers (QCLs), is particularly challenging. Due to the fast intersubband carrier transitions, single mode THz QCLs resemble Class A lasers, where chaos can be hardly excited, even with external perturbations. In this work, we experimentally show a THz chaos source based on a sole multimode THz QCL without any external perturbations. Such a dynamical regime is characterized by the largest Lyapunov exponent associated to the temporal traces of the measured radio frequency (intermode beatnote) signal of the laser. The experimental results and chaos validation are confirmed by simulations of our model based on effective semiconductor Maxwell-Bloch Equations. To further understand the physical mechanism of the chaos generation in THz QCLs, a reduced model based on two coupled complex Ginzburg-Landau equations is derived from the full model cited above to systematically investigate the effects of the linewidth enhancement factor and group velocity dispersion on the chaotic regime. This model allows us to show that the chaos generation in the THz QCL can be ascribed to the system attaining the defect mediated turbulence regime. Our findings pave the way for the generation of controllable and integrated THz chaos sources, as well as potential applications.

\end{abstract}

Chaos characterized by aperiodic deterministic dynamics in nonlinear systems has played a pivotal role in interpreting and controlling various ordered and disordered behaviors since its discovery\cite{lorenz1963}. In 1975, the principle of chaos generation in a laser system was first proposed\cite{HAKEN1975}. Subsequently, the chaos phenomena were observed in several types of lasers, e.g., gas lasers\cite{Weiss1988}, solid-state lasers\cite{Bracikowski1991}, fiber lasers\cite{Gregory1998}, and semiconductor lasers\cite{Deng2022}. Among the different kinds of lasers, semiconductor lasers serve as an ideal platform for the chaos investigation owing to its all solid-state nature, on-chip integration, and rich nonlinearities. Recently, extensive studies have been carried out on chaos phenomena in near-infrared semiconductor lasers, driven by their potential applications in high bit-rate optical communications\cite{Spitz2021}, random number generations\cite{Uchida2008}, light detection and ranging systems (LIDAR)\cite{Chen2023,Kanter2010,Kazuyuki2012}, etc. However, in the terahertz (THz) frequency range, due to the lack of effective THz light sources and high-speed detectors, chaos generation in THz semiconductor lasers is very challenging. The only THz chaotic behaviour was experimentally observed in superconducting Josephson junctions\cite{gulevich2019bridging} which required heavy cryogenic cooling down to 4.2 K and were not compatible with other III-V semiconductor devices. Therefore, it is urgent to develop semiconductor laser based THz chaotic sources for various applications as mentioned above. For instance, the THz chaotic LIDAR system can offer higher spatial resolution and greater resilience to interference\cite{Chen2023}; Chaos-based broadband THz spectroscopy can be employed to investigate physical and chemical processes in biological and living systems, including transient biological structures, unstable molecules, and chemical reactions, which is challenging to observe using conventional spectroscopy tools\cite{gulevich2019bridging}.

Chaos generation in lasers are strongly dependent on the type of the laser. Lasers are typically classified into three classes—Class A, B, and C—based on the relations among the decay rates of the carriers, polarization and photons\cite{arecchi1984deterministic,Virte2013,Sciamanna2015}. Most semiconductor lasers belong to Class B laser systems, being the carrier lifetime much longer than the photon lifetime\cite{ohtsubo2017}. Examples include quantum well lasers\cite{simpson1994}, quantum dot lasers\cite{kreinberg2019}, inter-band cascade lasers\cite{Deng2022}, and so on. Chaos generation in holonomous, single mode Class B lasers is ruled out by system dimensionality, unless a time-dependent external perturbation is introduced\cite{Maruskin2018}, such as optical feedback\cite{Mukai1985}, optical injection\cite{WIECZOREK20051}, optoelectronic feedback\cite{Tang2001}, loss modulation using a saturable absorber\cite{Viktorov2007}, etc. However, when a multimode semiconductor laser is considered, it has been shown that a sufficiently strong nonlinear mode coupling between several longitudinal or transverse modes of semiconductor lasers may induce chaotic instabilities even in the absence of additional parameter perturbations\cite{Sciamanna2015}. For example, chaos was observed in free-running vertical-cavity surface-emitting lasers without external perturbations, due to the nonlinear coupling between two elliptically polarized modes\cite{Virte2013}. In addition, free-running quantum-dot micropillar lasers exhibited chaos when operated close to the quantum limit\cite{kreinberg2019}.

The semiconductor lasers mentioned above normally emit at near-infrared wavelengths. To further extend the operation wavelength to mid-infrared, far-infrared or THz wavelengths, the quantum cascade laser (QCL)\cite{Faist1994, Kohler2002} is one of the most suitable candidates. Different from interband Class B lasers, QCLs have a fast gain recovery time (on the order of 1-100 ps)\cite{derntl2018} due to the fast carrier intersubband transitions, and therefore are normally categorized as Class A lasers\cite{heckelmann2023,opavcak2019}. In common sense, it is hardly possible to achieve chaos operation in a single mode free-running Class A laser, unless one or several independent control parameters are added. In recent years, exploiting such control techniques, it was possible to achieve and study the chaos generation, route to chaos, and modulation of chaos in single mode mid-infrared QCLs\cite{Jumpertz2016,chen2021,Spitz2019}. These studies demonstrated that chaos in mid-infrared QCLs can be obtained and it emerges in the temporal nonlinear dynamics driving the evolution of both photon and carrier densities, by applying optical feedback or optical injection into the laser. It is worth noting that the QCL exhibits a moderate but non-zero linewidth enhancement factor (LEF, or $\alpha$ factor) and when operated in a Fabry-Perot resonator configuration, the fowrard and backward propagating fields give rise to spatial hole burning (SHB) effects\cite{piccardo2022}. Both such effects contribute to the transition from single- to multimode emission in QCLs with low threshold in terms of laser pump. When the nonlinear effects cooperate with the group velocity dispersion (GVD), two operational regimes can be observed in QCLs: one is the mode-locking or the frequency comb regime\cite{piccardo2019frequency,opavcak2019,Guan2023}; the other is the chaos state.

Here, we demonstrate the chaos generation in a free-running multimode THz QCL emitting around 4.2 THz, without any external perturbations. The intermode beatnote (BN) of the THz QCL is measured and analysed while changing the bias current. Single narrow BN, multi-peak BN, and broad BN spanning over several GHz (chaotic state) are experimentally observed at different current pump conditions. Furthermore, the largest Lyapunov exponents calculated from the measured time traces of the broad intermode BN signals clearly prove the chaotic behaviour. To further prove the chaos generation in the free-running THz QCL, a full model based on the Maxwell-Bloch equations and a reduced model based on two coupled complex Ginzburg-Landau equations are, respectively, employed to reproduce the experimental data and unveil the physical mechanism for the chaos generation in THz QCLs.

\section*{Results}

\section*{Experimental setup and laser performance}

\begin{figure}[!t]
\centering
\includegraphics[width=0.98\linewidth]{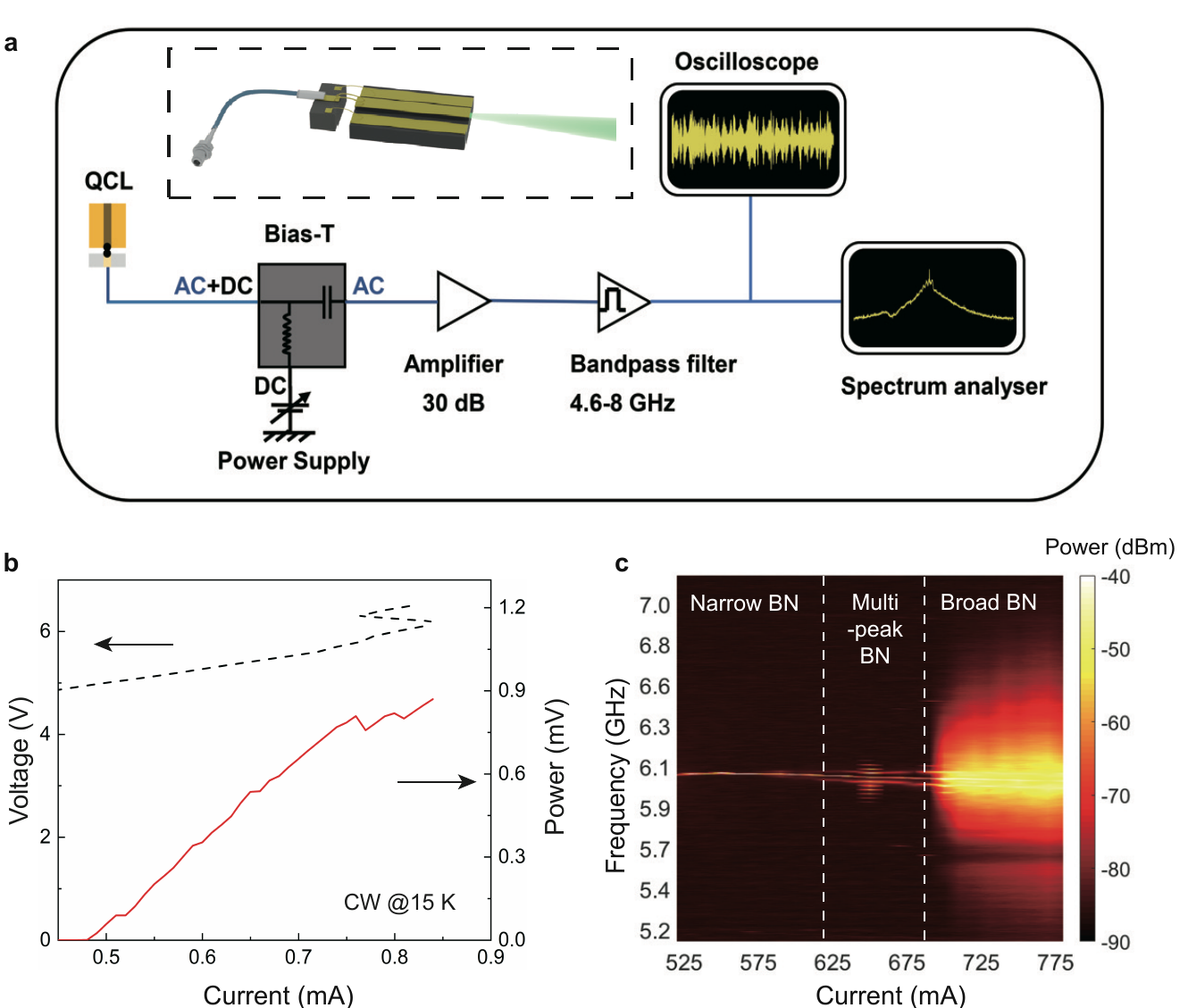}
\caption{Experimental setup and laser performances. (\textbf{a}) Experimental setup employed for chaos characterizations of THz QCLs. The laser self-detection scheme is used for the measurement of intermode BN signals which are finally recorded in frequency and time domains using a spectrum analyzer and a high speed oscilloscope (20 GS/s), respectively. To extract cleaner intermode BN signals, a microwave amplifier with a gain of 30 dB and a bandpass filter (4.6-8 GHz) are used. (\textbf{b}) Light–current–voltage characteristics of the QCL measured in cw mode at 15 K. (\textbf{c}) Intermode BN maps of the QCL, measured with resolution bandwidth (RBW) of 10 kHz and video bandwidth (VBW) of 1 kHz. The current step is 10 mA. The vertical dashed lines identify three current intervals corresponding to different operating regimes.}
\label{setup}
\end{figure}

Figure \ref{setup}a shows the experimental setup employed for the chaos characterization for THz QCLs. As previously explained, the time and frequency characteristics of the laser are measured using the laser itself as a detector. A microstrip line which is mounted close to the THz QCL chip is used to achieve the impedance matching between the laser (or detector) and the external circuit for the transmission of high frequency signals. A bias-T is used to provide the DC current to the THz QCL for lasing and simultaneously the AC port of the bias-T is used to extract the high frequency components of the photocurrent signal. After the bias-T, the RF signal is amplified by 30 dB using a microwave amplifier. The intermode beatnote signals which are originated from the mixing of THz modes are finally recorded using a spectrum analyser (frequency domain) and a high speed oscilloscope (time domain) as shown in Figure \ref{setup}a. The inset of Figure \ref{setup}a shows the three-dimensional geometry of the THz QCL mounted with a microstrip line and high speed cable for high frequency extractions. The THz QCL used in the experiment has a cavity length of 6 mm and a ridge width of 100 $\mu$m. The design and fabrication of the THz QCL are detailed in Methods.

Analyzing the chaos behaviour primarily involves the time and frequency characteristics of a laser. For single mode QCLs, a photodetector that can convert the optical signal to electrical one is normally employed to characterize the time and frequency traces in the range from DC to hundred MHz\cite{Deng2022,Jumpertz2016}. Note that in the lower frequency range around DC, the noise is usually large. Therefore, in this work, for the chaos in multimode THz QCLs, we can alternatively measure the RF signal around the intermode BN frequency of the laser rather than the signal around DC. In principle, the frequency and time characteristics around the fundamental intermode BN are identical to those measured around DC. But, the cleaner signal can be obtained when we perform the measurement at higher frequencies around the intermode BN due to the lower noise. Moreover, to detect the intermode BN signal of the laser, the laser self-detection scheme is employed\cite{Li2015oe,li2022oe}. Due to the fast carrier relaxation time in THz QCLs, the QCL itself can be used as a fast mixer to detect the beating of the longitudinal modes in the QCL cavity. The beating of the modes will bring about the modulation of population inversion which is then converted into current modulation. The current modulation can be finally measured directly from the laser and displayed on a spectrum analyser or a fast oscilloscope.

Figure \ref{setup}b demonstrates the light–current–voltage ($L-I-V$) characteristics of the THz QCL measured in continuous wave (CW) mode at a heat sink temperature of 15 K. The laser shows a maximum output power of 0.9 mW and a threshold current of 470 mA at 15 K. Note that the power shown here is the collected power without considering any calibrations for water absorption, window transmission, light collection efficiency, and so on. The emission spectra of the QCL measured using a Fourier transform infrared (FTIR) spectrometer is shown in Figure S1, Supplementary Information. Figure \ref{setup}c shows the intermode BN map as a function of bias current measured when the laser is operated in free-running. Three current regions, labeled as narrow BN, multi-peak BN, and broad BN, can be clearly observed. When pumped at lower currents between 525 and 620 mA, single-peak narrow BNs at 6.08 GHz are obtained, indicating the frequency comb operation. From 620 to 690 mA, we can see a clear change of the intermode BN signal from single peak to multi-peak behaviour. As the current grows beyond 690 mA, the intermode BN signal is characterized by a broad bandwidth spanning over 1 GHz, which is typically associated to a chaos emission.

\section*{Chaos measurement and analysis}

\begin{figure}[!b]
\centering
\includegraphics[width=0.98\linewidth]{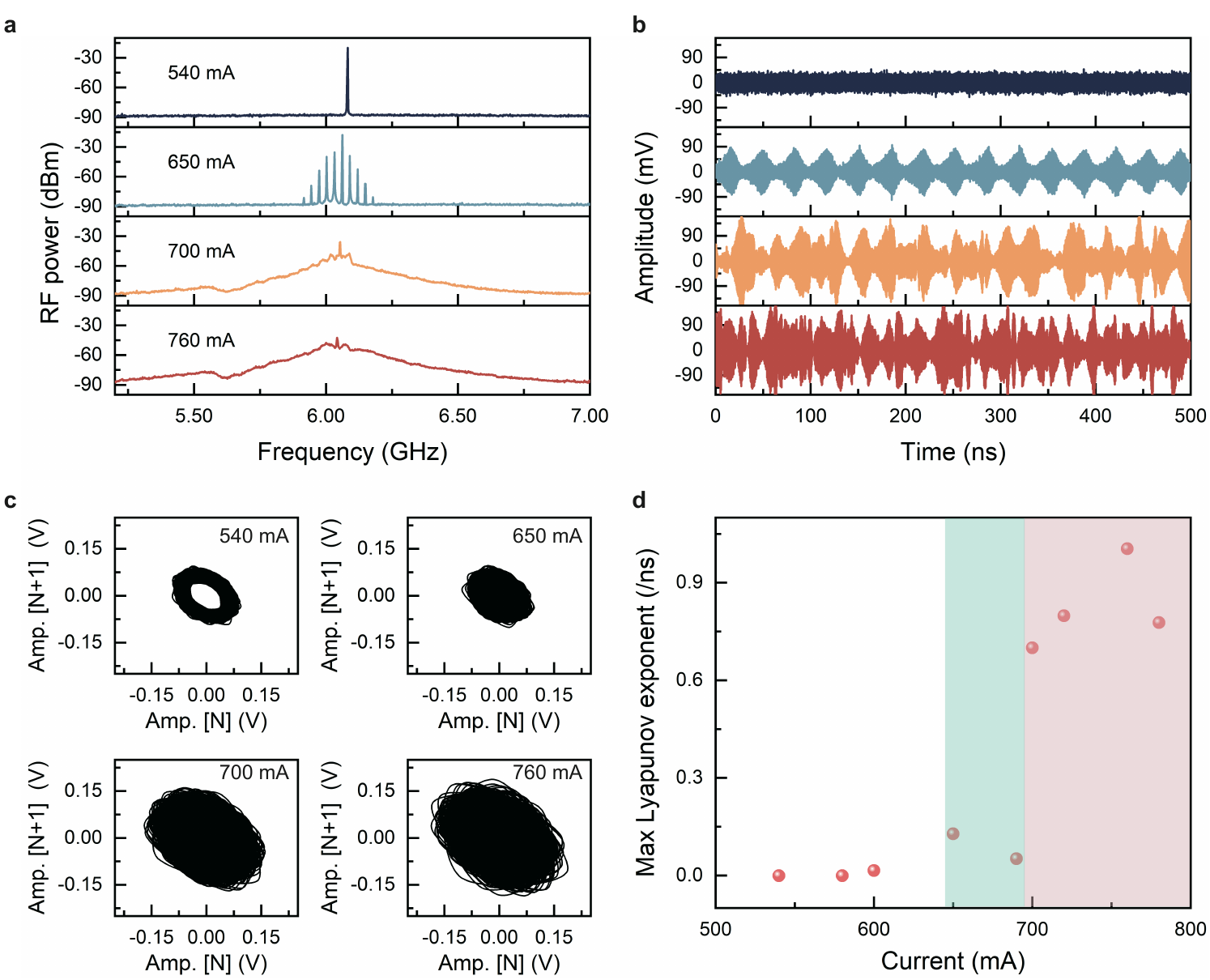}
\caption{Experimental results. RF spectra (\textbf{a}) and corresponding time traces (\textbf{b}) measured at different currents. The RF spectra were recorded with RBW of 10 kHz and VBW of 1 kHz. In \textbf{a} and \textbf{b}, from top and bottom panels, the current is increased from 540 to 760 mA. (\textbf{c}) Phase portraits at various currents obtained from time traces with a time length of 500 ns. (\textbf{d}) Calculated maximum Lyapunov exponents. The shaded pink region corresponds to the chaotic regime and the shaded green region represents the transition region from comb to chaos.}
\label{exp}
\end{figure}

In Figs. \ref{exp}a and \ref{exp}b, we show the recorded frequency spectra and corresponding time traces of the intermode BN signals, respectively, measured at different bias currents. As the current is raised from 540 to 760 mA (from top to bottom panel in Fig. \ref{exp}a), intermode BN spectra, recorded with a resolution bandwidth (RBW) of 10 kHz and a video bandwidth (VBW) of 1 kHz, show clear transitions from single narrow peak (frequency comb), multi-peak, to broad peak. The corresponding time traces of the signal are shown in Fig. \ref{exp}b for a time duration of 500 ns. At 540 mA, as shown in the top panel of Fig. \ref{exp}b, the time trace exhibits a sinusoidal periodic time series. As the current is increased to 650 mA, a modulation on the sinusoidal time trace is clearly observed which corresponds to the multi-peak intermode BN. When we further increase the current to 700 or 760 mA, typical chaotic time traces without any accurate and predictable periodicity are observed showing the laser enters the chaotic regime. Additionally, the frequency and temporal characteristics of BN at more current values are shown in Figures S2 and S3, Supplementary Information. Note that even in the chaotic regime, the fundamental oscillations around intermode BN frequencies can be always observed. These fundamental oscillations exist in all time traces measured at different currents, see Figure S4 in Supplementary Information.

The phase portrait of a time series can be constructed by applying a time lag on the original time series\cite{renjini2020phase}. For example, here single time step ($\tau$) employed in the time trace measurements is used as the time lag. In Fig. \ref{exp}c, four phase portraits calculated from time traces shown in Fig. \ref{exp}b at different currents are plotted. A periodic system normally shows an inner circle in its two-dimensional phase portrait, as shown in the first panel of Fig. \ref{exp}c for 540 mA. At 650 mA, the phase portrait becomes irregular, and the inner circle disappears. The phase portraits become more complex at higher currents, e.g., 700 and 760 mA, and much stronger oscillations result in larger trajectories in phase portraits. Furthermore, Video 1 in the Supplementary Information visually shows the trajectories for the four different cases mentioned above. At 540 mA, it can be clearly seen that a regular winding around the origin is obtained, which indicates a period-one orbit (comb operation). As the current is increased, the system demonstrates a more complex behaviour, i.e., the trajectory is firstly refined in an inner zone loop, followed by an eccentric shift toward the outer region. Note that at 650 mA, although the trajectory is much more complex than that obtained at 540 mA, a repeated amplitude-time trace can be observed, which refers to a quasi-periodic oscillation. The non-regular traces observed at 700 and 760 mA clearly indicate the chaotic state. We also re-plot the trajectories by reducing the data sampling rate, e.g., data were collected for every 1500$\tau$. The results summarized in Video 2 (Supplementary Information) show that at 540 mA the trajectory almost converges to a single point indicating a perfect periodic oscillation (comb state). Similar expansion of the trajectories with current is also observed in Video 2.

We offer a specific proof of chaos in our QCL emission by evaluating the Lyapunov exponents of the recorded series. We extract the largest Lyapunov exponent as shown in Fig. \ref{exp}d by using Wolf’s algorithm\cite{WOLF1985285}. Calculating the largest Lyapunov exponents has always been a powerful tool to identify chaos, reflecting the average exponential rate of divergence for nearby orbits in the phase space. In a stable system, the exponents are negative; while in a chaotic system, at least one positive Lyapunov exponent exists. From Fig. \ref{exp}d, we can see that as the current is below 600 mA, the largest Lyapunov exponents are negative values, which indicate stability with narrow intermode BN signals. When the QCL presents multi-peak BNs between 600 and 700 mA, the largest Lyapunov exponents become positive but smaller than 0.15/ns, which shows the transition region from comb to chaos. As broad intermode BNs appear ($>$700 mA), the calculated max Lyapunov exponents become much larger and especially it reaches 1/ns at 760 mA as shown in the shaded area in Fig. \ref{exp}d, which indicate the system enters the chaotic regime.

\section*{Simulation model and results}

\begin{figure}[!t]
\centering
\includegraphics[width=0.98\linewidth]{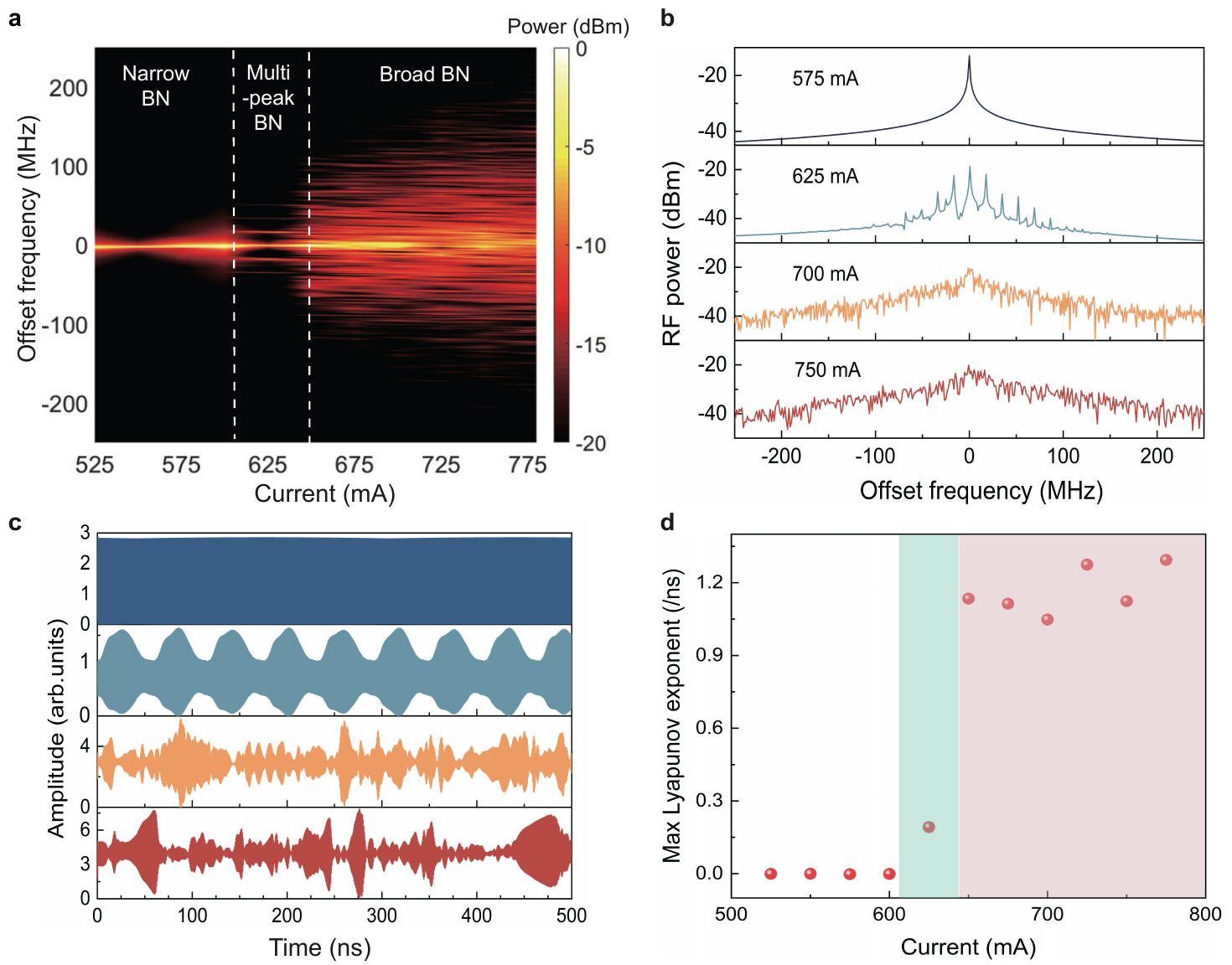}
\caption{Simulated results obtained from the full model. (\textbf{a}) Intermode BN map of the THz QCL. The current step is 25 mA. The vertical dashed lines identify three distinct regimes, i.e., single narrow BN, multi-peak BN, and broad BN. Simulated RF spectra (\textbf{b}) and corresponding time traces (\textbf{c}) for four different bias currents. In \textbf{a} and \textbf{b}, the frequencies have been offset by the center frequency of the intermode BN signal. In \textbf{b} and \textbf{c}, the drive current is increased from 575 to 750 mA from top to bottom panel. (\textbf{d}) Calculated largest Lyapunov exponents.}
\label{fullsim}
\end{figure}

We provide here sets of simulations to validate the chaotic characteristics of the QCL emission as observed in the experiment as shown in Fig.~\ref{exp}. The effective semiconductor Maxwell–Bloch equations (ESMBEs) have been adopted to investigate the nonlinear dynamics in semiconductor lasers\cite{columbo2007}, and specifically QCLs\cite{Columbo:18,Silvestri:20,silvestri2022multimode}. In this work, we thus presently adapt this model to the Fabry–Perot (FP) laser configuration to study the chaotic behavior in THz QCLs, tailoring it to the characteristics of the device studied in the experiments adopting the parameter set in Table S1, provided in the Supplementary Information. Figure~\ref{fullsim}a displays the simulated intermode BN map of the THz QCL. We observe that the sequence of regimes observed in Fig.~\ref{setup}c (single narrow BN, multi-peak BN, and broad BN) is reproduced with remarkable agreement in the simulations, although we note a slight discrepancy in the bias current values at which they occur. This is likely due to some additional dissipative effects or efficiency issues in the real device, which are not encompassed by the model. 

Figure \ref{fullsim}b shows four sections of the intermode BN map (Fig.~\ref{fullsim}a) obtained at 575, 625, 700, and 750 mA (from top to bottom panels). Note that to avoid the influence of higher harmonic frequency components on the time traces, and thus enable a direct comparison with the experimental traces (Fig. \ref{exp}b), a digital filtering method is implemented. The filtered simulated time traces are shown in Fig.~\ref{fullsim}c, while the original traces without filtering are shown in Figure S5 (Supplementary Information) for reference. We highlight that the traces in Fig.~\ref{fullsim}c represent the filtered output power retrieved from the ESMBE model, which can be compared with the voltage traces in Fig. \ref{exp}b, given the proportionality between these two quantities\cite{Taimre15,Kane}. Similar to Fig. \ref{exp}b, the simulated traces show clear evolution from a pure sinusoidal oscillation at 575 mA, to a modulated sinusoidal time trace at 625 mA, and finally reach the chaotic regime beyond 700 mA. To see the detailed structure of the time traces, we show the zoom-in of Fig. \ref{fullsim}c in a time scale of 20 ns in Figure S6 (Supplementary Information).

Consistently with what was done for the experimental traces, the Lyapunov exponents have also been computed for the simulated regimes. In Fig.~\ref{fullsim}d, the calculated largest Lyapunov exponents are plotted as a function of the bias current for the simulated traces. We observe the same transition to the chaotic regime as that observed in the experiment (see Fig.~\ref{exp}d). This is an excellent evidence of the concordance between the experiment and the model. In detail, we find that below 600 mA, the largest Lyapunov exponents are negative, an indication of regular laser dynamics. As the current is increased to 625 mA, an intermediate regime (green band) with exponents around 0.17/ns is observed. Beyond 650 mA, the values of the Lyapunov exponents range between 0.9 and 1.3/ns (see the shaded pink region in Fig. \ref{fullsim}d). This region is therefore characterized by chaotic dynamics, with Lyapunov exponent values of the same order of magnitude as those calculated for the experimental traces in the chaotic region (see the shaded area in Fig.~\ref{exp}d). To summarize, the simulations based on the full model clearly demonstrate a strong agreement with the experimental findings. This agreement is evident both in the dynamic scenario as the bias current varies, where we observe the onset of chaos transitioning through narrow BN and multi-peak BN regimes, and in the similarity between the experimental and simulated time traces and RF spectra. Furthermore, there is also consistency between simulation and experiment regarding the order of magnitude of the Lyapunov exponents within the chaotic region.

\begin{figure}[!t]
\centering
\includegraphics[width=0.65\linewidth]{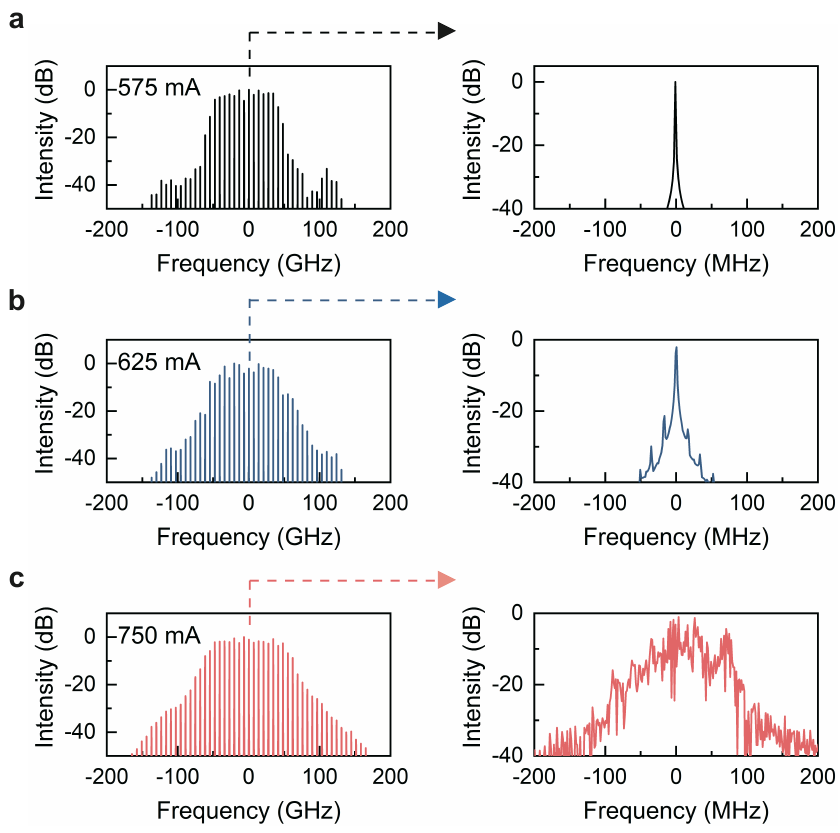}
\caption{Simulated optical spectra and the linewidth analysis. Left column: Optical spectra at 575 (\textbf{a}), 625 (\textbf{b}), 750 mA (\textbf{c}). Right column: zoom around one peak of the optical spectrum in the left column. For each spectrum, the intensity values are normalized with respect to the maximum.}
\label{opt}
\end{figure}

Since the simulated sequences effectively replicate the experimental ones, this reassures us about the transition to the chaotic regime, characterized by a shift from narrow to broad linewidths. In principle, when the laser enters a chaotic state, the frequency fluctuations increase dramatically, resulting in a significant broadening of the spectral linewidth, which is caused by the collapse of coherence\cite{dente1988chaos,WIECZOREK20051,ohtsubo2017}. However, experimentally characterizing the linewidth of optical modes in the THz frequency range poses challenges due to the limited spectral resolution of THz spectrometers (see the optical spectra of the THz QCL in Figure S1, Supplementary Information). Here, we analyze the optical lines in the simulated traces to determine if optical chaos is indeed present. In the left column of Fig. \ref{opt}, the simulated optical spectra at three bias current values corresponding to three different dynamical regimes are shown: 575 mA (comb operation with a single narrow BN), 625 mA (multi-peak BN), and 750 mA (broad BN). In the right column, a zoom of a single optical line from the corresponding spectrum shown in the left column is displayed. We observe that at 575 mA, the single optical line of the frequency comb exhibits a narrow linewidth (Fig. \ref{opt}a). Increasing the current to 625 mA (Fig. \ref{opt}b), the zoomed-in view of the single optical line reveals multiple peaks, which result in the multi-peak feature in the intermode BN (see Figs. \ref{exp}a and \ref{fullsim}b). As the current is increased to 750 mA, where the max Lyapunov exponent is large and positive, flagging chaotic regime, it can be seen that the optical spectrum is significantly broadened and simultaneously each optical line also demonstrates broad linewidth which spans over 200 MHz (Fig. \ref{opt}c). The chaos characterized by positive Lyapunov exponent in regimes of broad BN in the RF range can thus be ascribed to a multimode chaotic dynamics. 


To gain deeper insights into the physical mechanisms behind chaos generation in THz QCLs, we developed a reduced model, based on the formerly described full ESMBE model, which under appropriate approximations such as the near-threshold operation, can be reduced to a pair of coupled complex Ginzburg–Landau equations (CGLEs)\cite{silvestri2024unified}:
\begin{eqnarray}
\frac{\partial F^+}{\partial \eta} + \frac{\partial F^+}{\partial t} &=& -F^+ + (1+i\alpha)\mu F^+ - (1+i\alpha)^2 (|F^+|^2 + 2|F^-|^2) F^+ \nonumber \\
&& + \left( \frac{\sigma^2}{\Gamma^2 (1+i\alpha)} + i \frac{v_g}{2\tau_p} k'' \right) \frac{\partial^2 F^+}{\partial \eta^2}\label{modrid1}
\end{eqnarray}
\begin{eqnarray}
-\frac{\partial F^-}{\partial \eta} + \frac{\partial F^-}{\partial t} &=& -F^- + (1+i\alpha)\mu F^- - (1+i\alpha)^2 (|F^-|^2 + 2|F^+|^2) F^- \nonumber \\
&& + \left( \frac{\sigma^2}{\Gamma^2 (1+i\alpha)}  + i \frac{v_g}{2\tau_p} k'' \right) \frac{\partial^2 F^-}{\partial \eta^2}\label{modrid2}
\end{eqnarray}
where $F^+$ and $F^-$ are respectively the scaled forward and backward electric field envelopes, $\alpha$ is the linewidth enhancement factor, $\mu$ is the scaled pump parameter, $\sigma$ is the ratio between polarization dephasing time $\tau_d$ and photon lifetime $\tau_p$, $\Gamma$ is the scaled gain bandwidth of the QCL medium, $\eta$ and $t$ are the scaled spatial and temporal coordinates, $v_g$ is the group velocity, and $k''$ is the waveguide GVD.

The reduced mathematical complexity of this model compared to the full ESMBEs, along with its ability to describe QCL dynamics in terms of a universal class of equations like the CGLEs, makes it better suited for obtaining insights into the origin of physical phenomena, such as the formation of chaos in our case. A description of the reduced model can be found in the section Methods. For further details, see ref. [\citenum{silvestri2024unified}]. We firstly exploit the reduced model to perform emission characterization at different values of the normalized pump parameter $p$, defined as the ratio between $\mu$ and its threshold value ($\mu_{\rm{th}}$), i.e., $p$=$\frac{\mu}{\mu_{\rm{th}}}$. Also in this case, we adopt parameter values consistent with our experimental device (see Table S2 in the Supplementary information). As $p$ increases, we observe a transition from single-mode emission ($p=1.01$) to a frequency comb ($p=1.9$), and finally to chaotic dynamics at $p=3.15$, which is shown in Figure S7 (Supplementary Information). The chaotic nature of this regime is rigorously assessed by calculating the Lyapunov exponent value, which is 1.16/ns. Additionally, in line with expectations, the other two regimes exhibit negative Lyapunov exponents, namely $-9.5\times10^{-4}$/ns and $-4.29\times10^{-3}$/ns, in agreement with the experimental results of Figs.~\ref{setup}c, \ref{exp}b, and S1 (Supplementary Information). However, we observe a slight discrepancy in the pump value associated with chaotic dynamics compared to both the experiment and the full model simulations. We attribute this difference to the approximations implemented for the derivation of the reduced model, particularly the assumption of near-threshold operation\cite{silvestri2024unified}. 

Having verified that the reduced model correctly reproduces the transition between dynamic regimes reported in the experiment as the pump varies, we can employ it for a more in-depth analysis of the effect of certain physical parameters on the formation of chaos.
Among all the physical parameters of QCLs, it has been highlighted in the literature that non-zero $\alpha$ factor and GVD have a significant impact on the multimode dynamics of QCLs, particularly on the formation of frequency combs\cite{piccardo2022,opavcak2019,SilvestriReview,NaturePiccardo}. Here, we examine the role played by these two parameters in the formation of multimode chaos in these devices.

\begin{figure}[p]
\centering
\includegraphics[width=0.98\linewidth]{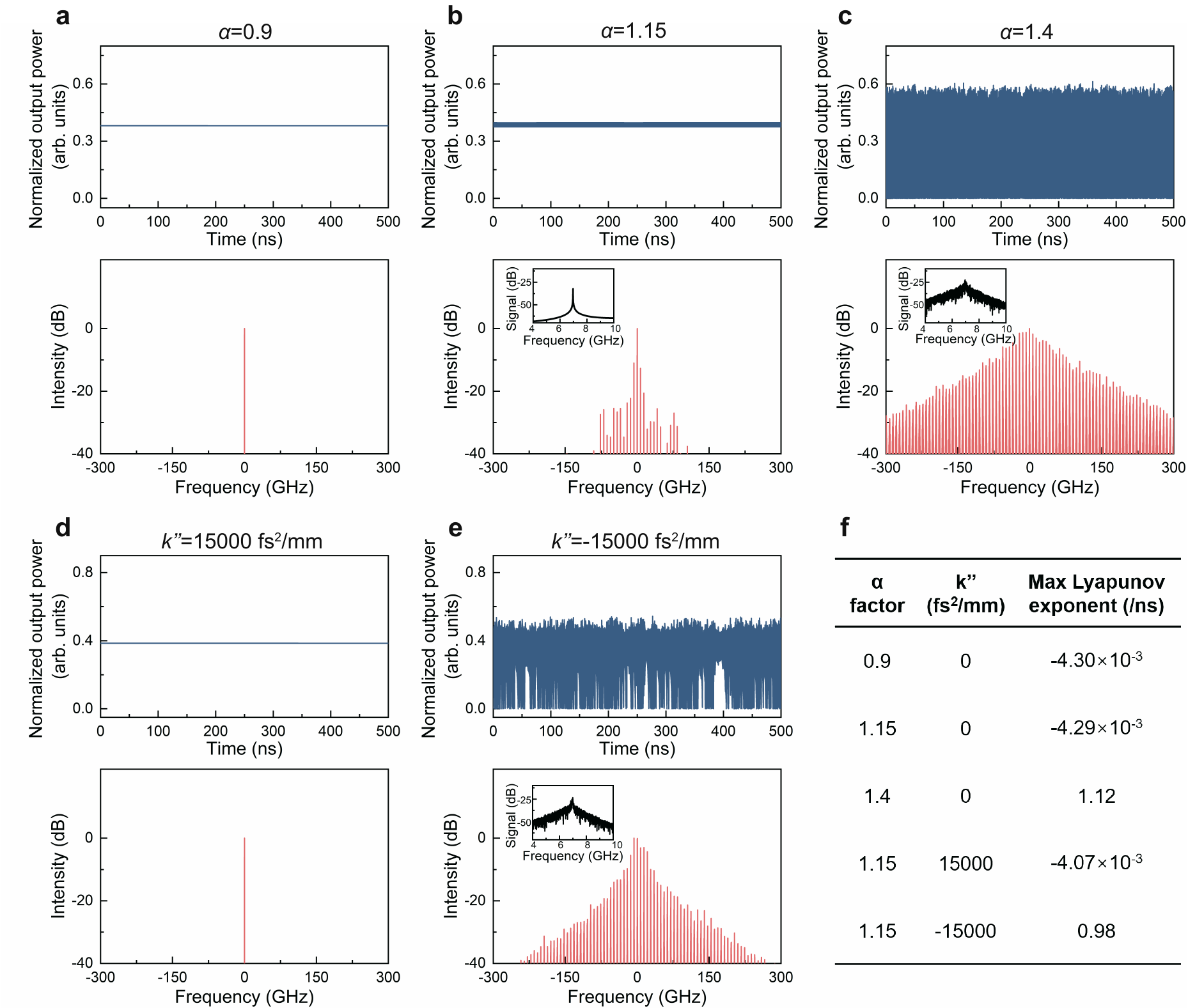}
\caption{Effects of $\alpha$ factor and GVD on the time and spectral characteristics of the THz QCL. (\textbf{a}),(\textbf{b}),(\textbf{c}), Scan of $\alpha$ factor from 0.9 to 1.4. The upper panels display the time traces by varying $\alpha$ factor and the corresponding optical spectra are shown in the lower panels. The insets in \textbf{b} and \textbf{c} are intermode BN spectra. For these simulations, the pump parameter is set to $p=1.9$ and the waveguide GVD is set to $k^{\prime\prime}=0$. (\textbf{d}),(\textbf{e}), Scan of $k^{\prime\prime}$. The upper panels show the time traces by varying $k^{\prime\prime}$ from -15000 to 15000 fs$^2$/mm and the corresponding optical spectra are shown in lower panels. The inset in \textbf{e} is the intermode BN spectrum. For these simulations, $p=1.9$ and $\alpha=1.15$. (\textbf{f}), Calculated largest Lyapunov exponents.}
\label{Reducedmodel}
\end{figure}

We first investigate the impact of the $\alpha$ factor on the temporal and spectral characteristics of THz QCL emission. To assess its effect, we vary $\alpha$ while keeping the values of the other parameters fixed. In addition to the parameter values provided in Table S2, we specify that the pump is set to $p=1.9$, while the GVD consists of the intrinsic material contribution, $\mathrm{GVD}_\mathrm{mat}=-77024~\mathrm{fs^2/mm}$. The waveguide GVD, denoted as $k^{\prime\prime}$, which is included as a phenomenological parameter in the model and can be varied, is set to zero. Details on the estimation of $\mathrm{GVD}_\mathrm{mat}$ and the incorporation of $k^{\prime\prime}$ into the model, which follows the approach used in ref.~[\citenum{columbo2021}], are provided in Methods. 
The values of $\alpha$ used in this study are consistent with experimental measurements reported in the literature for this parameter in QCLs\cite{SilvestriReview,piccardo2022,Grillot16}. The results are depicted in Figs.~\ref{Reducedmodel}a-\ref{Reducedmodel}c. We note that an increase in $\alpha$ from 0.9 to 1.15 results in the destabilization of the single-mode emission of Fig.~\ref{Reducedmodel}a in favor of a multimode state, in this case a frequency comb (Fig.~\ref{Reducedmodel}b). This is in line with expectations, as the increase in $\alpha$ corresponds to an increase in the coupling between phase and amplitude of the field, identified as one of the mechanisms responsible for the destabilization of CW emission in QCLs\cite{NaturePiccardo,Columbo:18}. A further increase in $\alpha$ to 1.4 leads to the emergence of an unlocked multimode state characterized by erratic behavior of the output power, and broadening of optical spectrum and fundamental intermode BN (Fig. \ref{Reducedmodel}c). 
As presented in the table of Fig.~\ref{Reducedmodel}f, the calculated Lyapunov exponent value of 1.12/ns for this regime confirms its chaotic nature, in contrast to the negative values found for the CW and comb states of Figs.~\ref{Reducedmodel}a-\ref{Reducedmodel}b. We note that the observed positive value of the Lyapunov exponent is very close to those reported for chaotic states in experiments and simulations with the full model (see Figs.~\ref{exp}d and \ref{fullsim}d). Furthermore, we highlight that theoretical studies have shown that an increase in the $\alpha$ factor, and thus in phase-amplitude coupling, leads to the emergence of irregularities in the dynamics of QCLs\cite{Silvestri:20,opavcak2019}, resulting in a loss of locking, identified by the broadening of the fundamental intermode BN in the RF spectrum. Here, through the calculation of the Lyapunov exponents, we demonstrate how this corresponds to the onset of chaos. Therefore, higher values of $\alpha$ favor the formation of chaos in free-running THz QCLs.


\begin{figure}
\centering
\includegraphics[width=0.82\linewidth]{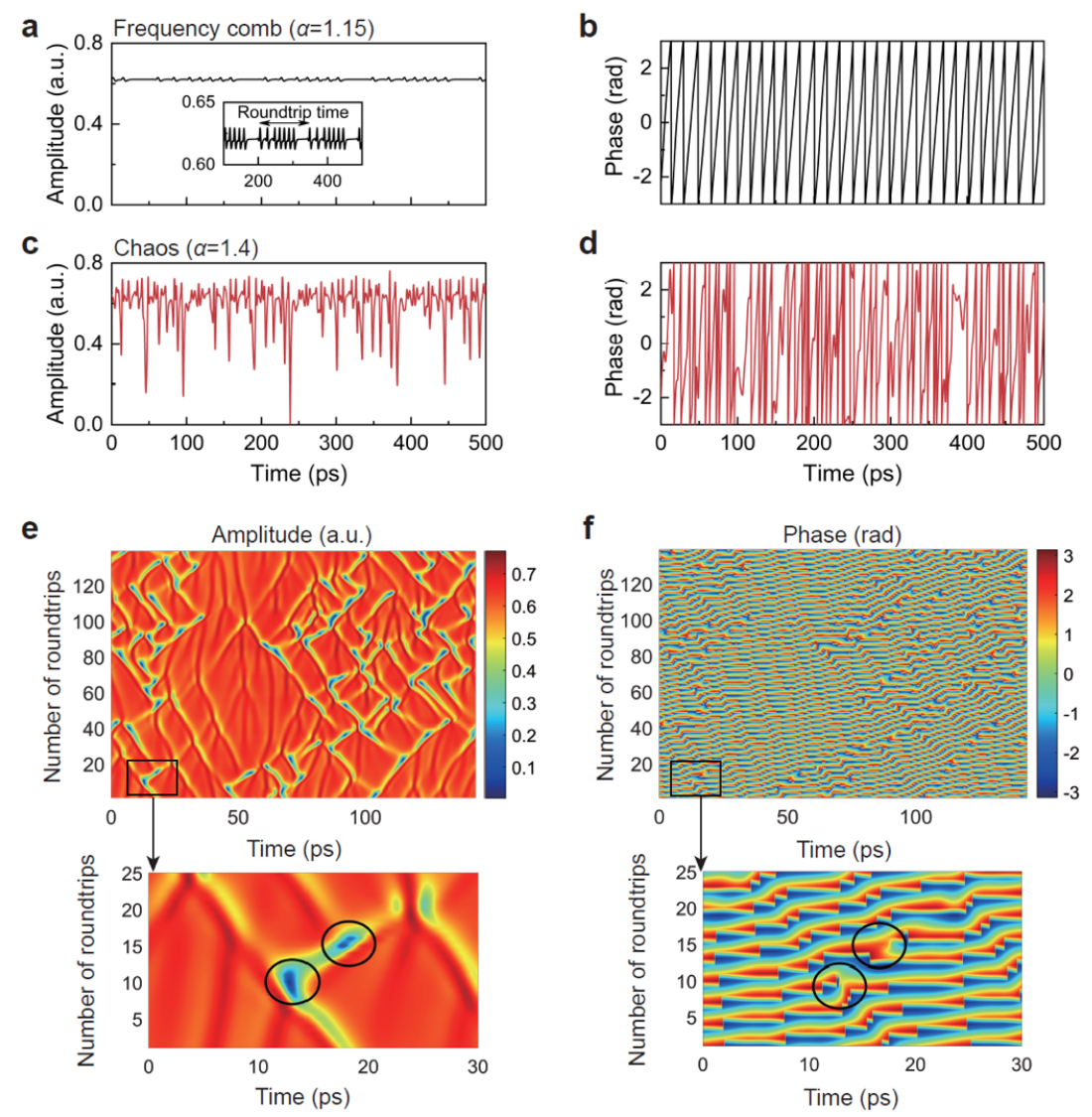}
\caption{Simulated amplitude and phase of the output field as a function of time for optical frequency comb (OFC) (\textbf{a},\textbf{b}), and multimode chaos regime (\textbf{c},\textbf{d}) of the THz QCL. For the OFC regime $\alpha$=1.15, while for the chaotic state $\alpha$=1.4. For both regimes $p$=1.9 and $k^{\prime\prime}$=0, and other parameters as in Table S2. The inset in (\textbf{a}) shows a zoom of the amplitude trace. 2D map for amplitude (\textbf{e}) and phase (\textbf{f}) of the chaotic regime of (\textbf{c},\textbf{d}) with time related to a roundtrip on the x-axis, and the number of roundtrips on the y-axis. The zoomed-in areas highlighted with a black rectangle are shown in the bottom panels, where defects in the field amplitude and the corresponding dislocations in the phase are circled in black. }
\label{defect}
\end{figure}

In addition to the $\alpha$ factor, we investigate the role of the GVD for chaos generation. Regarding comb operation in QCLs, the crucial role of the GVD is well known, since it has been shown that it regulates the field phase dynamics, contributing to establish a regime of chirped instantaneous frequency and thus determining the modulation frequency behavior of QCL combs\cite{opavcak2019}. Furthermore, the GVD required for comb operation depends on the $\alpha$ factor value. This implies that the range of GVD values where unlocked dynamics are observed features a dependence on $\alpha$ factor\cite{opavcak2019}. In our study, we fix $\alpha=1.15$ and investigate the formation of chaos as the GVD varies. The contribution of the QCL material, GVD$_\mathrm{mat}$, is fixed to -77024 fs$^2$/mm, while we tune $k^{\prime\prime}$ by setting it to -15000, 0, and 15000 fs$^2$/mm. Consequently, the total GVD ranges from -92024 fs$^2$/mm to -62024 fs$^2$/mm, values compatible with experimental measurements of this parameter for THz QCLs\cite{BachmannGVD}. In Figs. \ref{Reducedmodel}d, \ref{Reducedmodel}b, and \ref{Reducedmodel}e, we show the simulated optical spectra and time traces obtained with the reduced model for $k^{\prime\prime}$=15000 fs$^2$/mm, 0, and -15000 fs$^2$/mm, respectively. The pump value is $p=1.9$ and the other parameters are as in Table S2. We observe a CW-comb-chaos sequence as the GVD decreases, similar to what we observed with increasing $\alpha$ in Figs.~\ref{Reducedmodel}a-c. We remark that the chaotic nature of the regime in Fig.~\ref{Reducedmodel}e has been assessed by evaluating the maximum Lyapunov exponent, which exhibits a positive value of 0.98/ns (see Fig.~\ref{Reducedmodel}f). Various experimental methods, such as chirped mirrors \cite{Burghoff2014}, Gires-Tournois interferometer (GTI) mirrors \cite{Villares:16,Hillbrand:18,Lu2018}, and coupled cavities \cite{wang2017short}, have been employed to tune the GVD in QCLs, primarily with the intent of enhancing frequency comb stability. Our results suggest that these techniques could also be used to enhance chaos operation in these devices. Furthermore, we note that the observed transition from comb to chaos as the GVD decreases aligns with previous studies, which showed that below a certain GVD threshold, the QCL multimode dynamics become unlocked due to a significant reduction in the slope of the intermode phase spectrum\cite{opavcak2019}.

At this point we further employ the reduced model to gain insight into the physical origin of chaos in QCLs. Figures \ref{defect}a-\ref{defect}d show amplitude and phase of the output field as a function of time for the simulated regimes depicted in Figs. \ref{Reducedmodel}b and \ref{Reducedmodel}c, corresponding to two different values of $\alpha$. In particular, Figs. \ref{defect}a and \ref{defect}b are related to the comb emission presented in Fig. \ref{Reducedmodel}b, while Figs. \ref{defect}c and \ref{defect}d represent the chaotic state of Fig.~\ref{Reducedmodel}c. We observe that for the comb state, the amplitude trace exhibits a regular repetition of shallow structures with a period given by the cavity roundtrip, superimposed on a constant background. Additionally, the phase spans the entire interval from $-\pi$ to $\pi$, with a regular repetition of sawtooth-like structures. It is interesting to note how this corresponds to the regime of frequency combs induced by phase turbulence associated with the formation of homoclons\cite{Aranson} as recently reported in unidirectional ring QCLs\cite{NaturePiccardo,OpacakNature2024} and not yet experimentally observed in the FP configuration. In contrast, the chaotic state (see Figs. \ref{defect}c and \ref{defect}d) exhibits the coexistence of erratic behavior for both amplitude and phase, the former with random oscillations between 0 (field defects) and a maximum value, and the latter presenting irregular dynamics while spanning the interval from $-\pi$ to $\pi$. These features correspond to the so-called defect-mediated turbulence (DMT) regime, which is well-known as a dynamic behavior for CGLE systems\cite{Aranson}. 

To better characterize this regime, we collected 
20 ns time traces of the field phase and amplitude and after dividing each of them in segments of duration equal to the cavity roundtrip time we stacked these segments on top of each other obtaining a spatio-temporal representation of the 1D system dynamics shown in Figs. \ref{defect}e and \ref{defect}f; on the horizontal axis it is reported the fast temporal scale associated with the system evolution during a single roundtrip, while the vertical axis shows its evolution over a much slower temporal scale measured in units of roundtrips. In the amplitude map, multiple points of zero intensity (defects), depicted in dark blue, can be identified and, as previously mentioned, are characteristic of the DMT regime. A zoomed-in view of a region containing two defects is shown in the bottom panel of Fig. \ref{defect}e, with the defects circled. The bottom panel of Fig. \ref{defect}f shows the corresponding region in the phase map, where dislocations (also circled) align with the defect positions in the amplitude map as expected (see e.g. Fig. 5 in ref. [\citenum{Chate_1994}] and Fig. 8 in ref. [\citenum{Aranson}]). Therefore, having reduced the dynamics of QCLs to a system of two coupled CGLEs, reproducing DMT states with parameters compatible with the experimental device, and the fact that the values of the Lyapunov exponents calculated for these DMT states are very close to those calculated for the chaotic experimental traces, suggest that the origin of chaos in THz QCLs can be attributed to the DMT.


\section*{Discussion}
In summary, we have demonstrated chaos generation in free-running THz QCLs. Experimentally, by varying the bias current, the intermode BN signal showed clear transitions from single narrow BNs around the laser threshold to multi-peak BNs, and further into a chaotic regime characterized by broad BNs, spanning several GHz at high bias currents. The calculated largest Lyapunov exponents from the experimental time traces exhibited positive values at high currents, indicating chaos generation. To the best of our knowledge, this is the first demonstration of THz chaos generation in semiconductor lasers (previous THz chaos generation was only observed in superconductors\cite{gulevich2019bridging}). In contrast to chaos generation in single mode Class A and Class B lasers that normally require external perturbations, here we presented the THz chaos in free-running THz QCLs by employing the laser multimode dynamics. Furthermore, the self-detection scheme significantly simplified the experimental setup for THz chaos characterizations.

Two theoretical models were employed to validate these experimental results and investigate the physical origin of chaos generation in THz QCLs. With the first model, consisting of a full set of effective semiconductor Maxwell–Bloch equations (ESMBEs), we successfully reproduced temporal traces, frequency spectra, and the dynamic scenario as the bias current varied, observed in the experiments. Additionally, the ESMBEs revealed that chaotic behavior is characterized by a broadening of each line in the optical spectrum, as expected. The second model, comprising two coupled complex Ginzburg–Landau equations (CGLEs), features lower mathematical complexity and is thus more suitable for providing insight into the physical origin of chaos in THz QCLs. Using this reduced model, we first demonstrated that an increase in the $\alpha$ factor, which implies stronger phase-amplitude coupling of the electric field, enhances the likelihood of chaotic emission. Furthermore, we observe that a reduction in the total GVD, which can be experimentally controlled using techniques such as chirped mirrors or Gires-Tournois interferometer mirrors, leads to a transition from a comb regime to a chaotic one. This is in accordance with previous studies, which have shown that a decrease in GVD implies a decrease in the slope of the intermode phase spectrum \cite{opavcak2019}, corresponding to a loss of locking. These findings provide essential guidelines for designing sources or experimental setups for chaos generation in the THz region. Moreover, this second approach suggests that the physical origin of chaos in free-running THz QCLs might be attributed to the well-known phenomenon of defect-mediated turbulence, typical of extended nonlinear and dissipative systems described by CGLEs.

A systematic exploration of chaotic regimes in THz QCLs presents significant potential, bridging gaps in existing literature and advancing the comprehension of QCL's dynamics. The ability to manipulate and control chaos in the THz region opens unprecedented avenues for innovative technologies with higher performance and versatility. These technologies include broadband spectroscopy, sensing, and free-space communications. Furthermore, chaotic QCLs find applications in LIDAR systems, offering jamming-resistant, high-resolution sensing, or secure multi-channel communications involving chaos modulation for message encryption or synchronized chaos for message transmission. Each tooth in chaotic optical frequency combs holds the potential for random bit generation. Consequently, the novel phenomena detailed in our study offer promising prospects for driving advancements across a diverse range of technological domains in the THz region.

\section*{Methods}
\textbf{THz QCL.} The active region of the THz QCL used in this work is based on a hybrid design where bound-to-continuum transitions are exploited for the generation of terahertz photons, while fast longitudinal optical phonon scattering allows the fast depopulation of the lower laser state. The active region is designed for light emission at 4.2 THz. The detailed layer structure of the QCL active region is available in the ref. [\citenum{wan2017}]. The entire active region structure was grown by a molecular beam epitaxy system on a semi-insulating GaAs (100) substrate. The grown wafer was then processed into a single plasmon waveguide geometry with a ridge width of 100 $\mu$m. To improve the thermal management, the laser substrate was thinned down to 100 $\mu$m using grinding and polishing techniques. Various cavity lengths can be made by directly cleaving the laser ridges. The THz QCL used in the experiment has a nominal cavity length of 6 mm and a ridge width of 100 $\mu$m. 

The continuous wave (cw) output power of the THz QCLs is measured using a THz power meter (Ophir, 3A-P THz), with the lasers operated in constant current mode. To capture THz light emitted from the laser front facet, two parabolic mirrors are employed to collect and focus the light onto a THz power detector. Additionally, the beam path is purged with dry air to reduce water absorption. The emission spectra shown in Figure S1 are measured with a Fourier transform infrared (FTIR) spectrometer (Bruker, v80) with a spectral resolution of 0.1 cm$^{-1}$ (3 GHz).

\textbf{Lyapunov exponents.} Lyapunov exponents represent the average exponential rates of divergence or convergence of nearby orbits in phase space\cite{WOLF1985285,sprott2003chaos}. In a periodic system, the exponents are negative, but in a chaotic system, at least one positive Lyapunov exponent exists. Calculating Lyapunov exponents typically involves three steps: First of all, time delays are computed using the cross-correlation method to determine the dynamic properties of the data. Secondly, the false nearest neighbor method is employed to find an appropriate embedding dimension for effectively representing the data's phase space structure. Finally, by reconstructing the phase space and applying mathematical fitting techniques, Lyapunov exponents are computed to assess the system's chaotic behavior and sensitivity.

\textbf{Full model based on effective semiconductor Maxwell-Bloch equations.} The effective semiconductor Maxwell-Bloch equations (ESMBEs) used in this work describe the multimode dynamics of quantum cascade lasers (QCLs) in the Fabry-Perot (FP) configuration \cite{Silvestri:20}. This model encompasses several fundamental properties of semiconductor materials, including asymmetric gain and refractive index profiles in the frequency domain, a non-zero linewidth enhancement factor, and the dependence of the  optical susceptibility on the carrier density. Additionally, the model accounts for spatial hole burning (SHB), manifested as a carrier grating induced by the interference of the counterpropagating fields within the FP resonator. The dynamical variables in the set of ESMBEs are the forward and backward electric fields and polarization terms, the carrier density, and the carrier grating. The ESMBE parameters used in the simulations are listed in Table S1, Supplementary Information.

\textbf{Reduced model based on complex Ginzburg-Landau equations.} The reduced model is based on two coupled complex Ginzburg-Landau equations (CGLEs), Eqs.~(\ref{modrid1})-(\ref{modrid2}), derived from the full ESMBE model under the assumption of fast carriers and near-threshold operation\cite{silvestri2024unified,columbo2021}, using an approach analogous to that recently employed to describe Kerr solitons in driven and passive microresonators\cite{Cole2018}. This model describes the multimode dynamics of QCLs in the Fabry-Perot (FP) configuration. The dynamical variables of the two CGLEs are the forward and backward counterpropagating fields inside the FP cavity. The coupling between these fields is due to spatial hole burning (SHB), which is included in the model. Additionally, the CGLEs account for a non-zero linewidth enhancement factor and group velocity dispersion. The parameters used in the simulations with the reduced model are listed in Table S2, Supplementary Information. In order to estimate the material GVD, we rearrange Eq.~(\ref{modrid1}), obtaining:
\begin{eqnarray}
\frac{\partial F^+}{\partial \eta} + \frac{\partial F^+}{\partial t} &=& -F^+ + (1 + i\alpha)\mu F^+ - (1 + i\alpha)2(|F^+|^2 + 2|F^-|^2) F^+ \nonumber\\&+& \left(\frac{\sigma^2}{\Gamma^2 (1+\alpha^2)} + i \frac{v_g}{2\tau_p} \left( k'' - \frac{2\tau_p \alpha\sigma^2}{v_g \Gamma^2 (1+\alpha^2)} \right) \right) \frac{\partial^2 F^+}{\partial \eta^2} \label{modrid3}
\end{eqnarray}
We define the total group velocity dispersion as:
\begin{equation}
\text{GVD}_{\text{tot}} = k'' - \frac{2\tau_p \alpha\sigma^2}{v_g \Gamma^2 (1+\alpha^2)}
\end{equation}
Therefore, we observe that $\text{GVD}_{\text{tot}}$ is expressed as the sum of \(k''\) (waveguide GVD) and the material GVD given by:
\begin{equation}
\text{GVD}_{\text{mat}} = -\frac{2\tau_p \alpha\sigma^2}{v_g \Gamma^2 (1+\alpha^2)}
\end{equation}
We highlight the diffusive and dispersive role of the differential term on the right-hand side of Eq.~(\ref{modrid3}). We can estimate $\text{GVD}_{\text{mat}}$ by using the parameters exploited in the numerical simulations (see Table S2), obtaining $\text{GVD}_{\text{mat}} = -77024 \, \text{fs}^2/\text{mm}$.





\begin{addendum}
\item This work is supported by the National Science Fund for Distinguished Young Scholars (62325509), the Innovation Program for Quantum Science and Technology (2023ZD0301000), the National Natural Science Foundation of China (62235019, 61875220, 61927813, 61991430, 62035005, 62105351, 62275258, 62035014 and 62305364), Science and Technology Commission of Shanghai Municipality(21ZR1474600), the ``From 0 to 1" Innovation Program of the Chinese Academy of Sciences (ZDBS-LY-JSC009), and the CAS Project for Young Scientists in Basic Research (YSBR-069). M.B. and L.L.C. acknowledge the research funding from Italian Ministerial PRIN project ``MIRABILIS" (CUPM. D53D23002780006).

\item[Data availability] 
The authors state that data generated in this study are provided within the article and Supplementary Information files. All data are available from the corresponding author upon request.
\item[Code availability] 
All code used in this study is available from the corresponding author upon request.
\item[Competing financial Interests] The authors declare that they have no competing financial interests.


\end{addendum}

\section*{References}

\bibliography{ref}

\section*{Author Contributions}
B.B.L., C. S., and K.Z. contributed equally to this work.
H.L. conceived the study. Z.P.L., X.H.M., B.B.L., and W.J.W. fabricated the THz QCLs. B.B.L., K.Z., X.H.M., S.M.W., and H.L. performed experimental measurements. C.S., M.B., and L.C. carried out numerical simulations based on the full model and reduced model. Y.Z., J.S.P. and H.P.Z. performed the analysis of the Lyapunov exponents of the THz QCL. H.L., L.C., B.B.L., C.S., K.Z., J.S.P., H.P.Z., C.W., and Z.Z.Z. participated in the date analysis. B.B.L., H.L, C.S., L.C., and K.Z. wrote the manuscript with contributions from all authors. H.L., L.C., and H.P.Z. supervised the project.

\end{document}